\RequirePackage[2020-02-02]{latexrelease}
\documentclass[aps,pra,twocolumn,superscriptaddress,tightenlines, footinbib, longbibliography]{revtex4-2}
\usepackage{graphicx}
\usepackage{dcolumn}
\usepackage{bm}
\usepackage[colorlinks,linkcolor=blue,anchorcolor=blue,citecolor=blue,urlcolor=blue]{hyperref}
\usepackage{color}
\usepackage{amsmath}
\usepackage{amsthm}
\usepackage{amssymb}
\usepackage{mathrsfs}
\usepackage{array}
\usepackage{booktabs}
\usepackage{multirow}
\usepackage{siunitx}
\usepackage{chemformula}
\usepackage{colortbl}
\definecolor{darkgrey}{RGB}{216,208,199}
\definecolor{lightgrey}{RGB}{233,231,232}
\usepackage{easyReview} 
\usepackage{soulutf8} 

\begin{document}
	
	\title{Nonreciprocal slow or fast light in anti-$\mathcal{PT}$-symmetric optomechanics}
	\author{Meiyu Peng}
	\affiliation{Key Laboratory of Low-Dimensional Quantum Structures and Quantum Control of Ministry of Education,\\ Department of Physics and Synergetic Innovation Center for Quantum Effects and Applications,\\ Hunan Normal University, Changsha 410081, China}
	
	\author{Huilai Zhang}
	\affiliation{College of Physics and Electronic Engineering, Zhengzhou University of Light Industry, Zhengzhou 450000, China}
	\affiliation{Synergetic Innovation Academy for Quantum Science and Technology, Zhengzhou University of Light Industry, Zhengzhou 450002, China}
	
	\author{Qian Zhang}
	\affiliation{Key Laboratory of Low-Dimensional Quantum Structures and Quantum Control of Ministry of Education,\\ Department of Physics and Synergetic Innovation Center for Quantum Effects and Applications,\\ Hunan Normal University, Changsha 410081, China}
	
	\author{Tian-Xiang Lu}
	\affiliation{College of Physics and Electronic Information, Gannan Normal University, Ganzhou 341000, Jiangxi, China}
	
	\author{Imran M. Mirza}
	\affiliation{Department of Physics, Miami University, Oxford, OH 45056, USA}
	
	\author{Hui Jing}
	\email{jinghui73@foxmail.com}
	\affiliation{Key Laboratory of Low-Dimensional Quantum Structures and Quantum Control of Ministry of Education,\\ Department of Physics and Synergetic Innovation Center for Quantum Effects and Applications,\\ Hunan Normal University, Changsha 410081, China}
	\affiliation{Synergetic Innovation Academy for Quantum Science and Technology, Zhengzhou University of Light Industry, Zhengzhou 450002, China}
	
	\date{\today}
	
	\begin{abstract}
		Non-Hermitian systems with anti-parity-time ($\mathcal{APT}$) symmetry have revealed rich physics beyond conventional systems. Here, we study optomechanics in an $\mathcal{APT}$-symmetric spinning resonator and show that, by tuning the rotating speed to approach the exceptional point (EP) or the non-Hermitian spectral degeneracy, nonreciprocal light transmission with a high isolation ratio can be realized. Accompanying this process, nonreciprocal group delay or advance is also identified in the vicinity of EP. Our work sheds new light on manipulating laser propagation with optomechanical EP devices and, in a broader view, can be extended to explore a wide range of $\mathcal{APT}$-symmetric effects, such as $\mathcal{APT}$-symmetric phonon lasers, $\mathcal{APT}$-symmetric topological effects, and $\mathcal{APT}$-symmetric force sensing or accelerator.
	\end{abstract}
	\maketitle

	\section{Introduction}\label{Sec:Introduction}
	
	Recently, exotic and often counterintuitive effects emerging in non-Hermitian systems, especially those with Hamiltonians that are invariant under the combined operation of parity and time inversions, have attracted intense interests across natural sciences~\cite{Bender1998,bender2013,feng2017,ElGanainy2018,Konotop2016,Oezdemir2019,Zyablovsky2014,ashida2020,zhang2022dynamical}. In the parity-time-symmetric ($\mathcal{PT}$-symmetric) phase, these non-conservative Hamiltonians can still exhibit entirely real spectra, while by surpassing the exceptional point (EP), i.e., the non-Hermitian spectral degeneracy of the system~\cite{miri2019Exceptional,zhang2022dynamical}, radical changes of the properties of the systems can happen in the $\mathcal{PT}$-symmetry-broken phase~\cite{Bender1998}. In practice, $\mathcal{PT}$ symmetry has been observed in diverse systems such as optical microcavities~\cite{Chang2014,Peng2014,Zhang2018a,Xu2016,jing2014pt,Jing2015,Jing2017}, atomic systems~\cite{zhang2016,Li2019b}, and acoustic devices~\cite{Zhu2014,Fleury2015,ding2016}, and unique EP effects have been demonstrated~\cite{feng2017,ElGanainy2018,Konotop2016,Oezdemir2019,Zyablovsky2014,ashida2020,lu2015p,Liu2016}, such as loss-induced transparency~\cite{peng2014loss,Zhang2018}, single-mode lasing~\cite{Feng2014,Hodaei2016}, enhanced light-matter interactions~\cite{jing2014pt,Zhang2018a}, and non-reciprocal laser propagation~\cite{Chang2014,Peng2014}, to name only a few.
	
	Unlike $\mathcal{PT}$-symmetric systems usually requiring gain mediums~\cite{Oezdemir2019}, purely lossy systems with anti-$\mathcal{PT}$ ($\mathcal{APT}$) symmetry~\cite{Ge2013,Peng2016}, i.e., $\{\mathcal{PT},H_{\mathrm{\mathcal{APT}}}\}=0$, has also drawn much attention, due to their intriguing properties such as energy-difference conserving dynamics~\cite{park2021,Choi2018} and shorter-length chiral mode switch~\cite{Feng2022harnessing}. In experiments, $\mathcal{APT}$ symmetry and breaking have been observed in a wide range of systems, such as radiative plasmonics~\cite{yang2022radiative}, thermal or cold atoms~\cite{Peng2016,Jiang2019,Cao2020,ding2022,he2022}, electrical circuits~\cite{Choi2018,Stegmaier2021}, magnonic systems~\cite{Yang2020,Zhao2020}, optical waveguides or microcavities~\cite{Zhang2020,Wang2020,Zhang2019,Fan2020,Bergman2021}, and diffusive systems~\cite{Li2019a,Cao2020a}.  In a recent work, the possibility of breaking $\mathcal{APT}$ symmetry by spinning a resonator has been proposed, which is a purely optical system without considering the coupling of photons and phonons~\cite{Zhang2020a}. Based on this work, here we study the hybrid optomechanical effects in such an $\mathcal{APT}$-symmetric system.
	
	Specifically, we study the process of optomechanically induced transparency (OMIT) in an $\mathcal{APT}$-symmetric spinning resonator. OMIT has been observed in many different systems and led to important applications~\cite{Xiong2018,Liu2017,SafaviNaeini2011,Agarwal2010,Weis2010,Teufel2011C,Karuza2013,Jiao2018,Zhang2018,lu2017,lu2018,Jing2015,Lu2019Selective,jiao2016nonlinear,mirza2019optical,wen2022optomechanically}, ranging from optical communications~\cite{Zhou2013} and light storage~\cite{Fiore2011,Zhu2007} to weak force measurements~\cite{Jiang2014} and mechanical cooling~\cite{Ojanen2014}. In particular, the abnormal dispersion accompanied by the OMIT can alter the light group velocity in a radical way~\cite{SafaviNaeini2011}. Here we show that nonreciprocal light transmission with high isolation ratio can be realized by breaking the $\mathcal{APT}$ symmetry. More importantly, we find the group delay is also nonreciprocal in $\mathcal{APT}$-symmetry-broken regime. Our work provides a promising approach to manipulate light propagation in lossy devices and can stimulate more works on $\mathcal{APT}$-symmetric optomechanics~\cite{jing2014pt,grudinin2010,Zhang2018a,Xu2016,zhao2020weak,anetsberger2009,Gavartin2012,Liu2016,Wiersig2020,Hodaei2017,Chen2017}. In comparison to the very recent work using a purely optical system~\cite{Zhang2020a}, here we focus on the hybrid optomechanical interaction in an $\mathcal{APT}$ symmetric system and, for the first time as far as we know, reveal the possibility of observing nonreciprocal slow or fast light by breaking the $\mathcal{APT}$ symmetry.
	
	\begin{figure*}[t]
		\centering
		\includegraphics[width=0.95\textwidth]{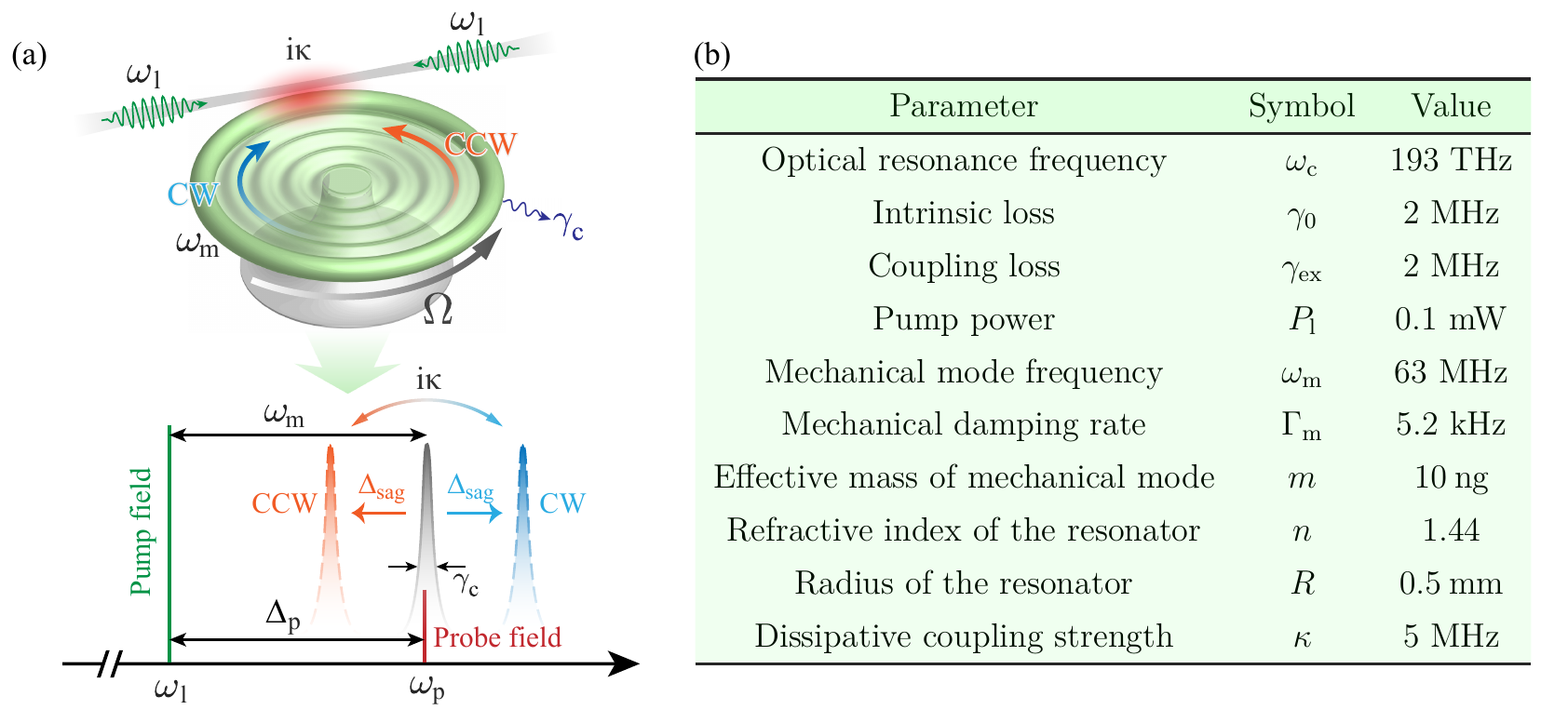}
		\caption{$\mathcal{APT}$-symmetric optomechanics based on a spinning resonator.
			(a) Schematic diagram and frequency spectrum of the system. The resonator is driven bidirectionally by pump fields with the same frequency $ \omega_{\mathrm{l}}$. The spinning of the resonator induces different Sagnac-Fizeau shift $ \Delta_{\mathrm{sag}} $ in CW and CCW modes of the resonator.
			(b) A summary of the experimentally accessible parameters, used for the numerical simulations~\cite{Maayani2018,righini2011}. \label{fig:figure1}}
	\end{figure*}	
	
	The paper is organized as follows. In Sec.~\ref{Sec:ModelandHamiltonian}, we introduce the model of our $\mathcal{APT}$-symmetric optomechanical system and calculate the optical transmission of such a system. Detailed discussions of the nonreciprocal effects of both the optical transmission and the group delay or advance in an $\mathcal{APT}$-symmetric spinning resonator are given in Sec.~\ref{Sec:Results1}, and finally we conclude in Sec.~\ref{Sec:Conclusions}.

	\section{$\mathcal{APT}$-symmetric optomechanics}\label{Sec:ModelandHamiltonian}
	
	We start by considering an optomechanical resonator which supports two counter-propagating optical whispering gallery modes (WGMs) and a mechanical breathing mode [see Fig.~\ref{fig:figure1}(a)]. A tapered-fiber waveguide is coupled to the resonator, which can give rise to the dissipative coupling between the optical WGMs also~\cite{Lai2019}. We note that a spinning resonator with a stationary waveguide was already experimentally realized~\cite{Maayani2018}, and in that experiment, the resonator with the radius $R=1.1\:\mathrm{mm}$ can spin with the stability of its axis, reaching the maximal frequency of $6.6\:\mathrm{kHz}$. Due to the Sagnac effect, the clockwise (CW) and counterclockwise (CCW) modes experience different Sagnac-Fizeau shifts~\cite{Malykin2000,Maayani2018}: $\omega_{\mathrm{c}} \rightarrow \omega_{\mathrm{c}} \pm \Delta_{\mathrm{sag}}$, where
	\begin{align}
		\Delta_{\mathrm{sag}}=\frac{nR\Omega\omega_{\mathrm{c}}}{c}\left(1-\frac{1}{n^{2}}-\frac{\lambda}{n}\frac{\mathrm{d}n}{\mathrm{d}\lambda}\right),
	\end{align}
	with $c$ ($ \lambda $) being the speed (wavelength) of light. $ n $ and $ R $ are the refractive index and radius of the resonator, respectively. $\omega_{\mathrm{c}}$ is the resonant frequency of the optical modes. $\Omega$ is the spinning speed and the dispersion term $ \mathrm{d}n/\mathrm{d}\lambda $, characterizing the relativistic origin of the Sagnac effect, is small in typical materials (up to about 1\%)~\cite{Malykin2000}.
	
	By fixing the CCW spinning direction of the resonator, in a frame rotating at a pump frequency  $\omega_{\mathrm{l}}$, the effective Hamiltonian of the optical part can be written as
	\begin{align}\label{EqOpticalHamiltonian}
		H_{\mathrm{0}}=\left(\begin{array}{cc}
			\Delta_{\mathrm{+}}-i\gamma_{\mathrm{c}} & i\kappa\\
			i\kappa & \Delta_{\mathrm{-}}-i\gamma_{\mathrm{c}}
		\end{array}\right),
	\end{align}
	where
	\begin{align}
		\Delta_{\pm}=\Delta_{\mathrm{c}}\pm\Delta_{\mathrm{sag}}, \quad \Delta_{\mathrm{c}}=\omega_{\mathrm{c}}-\omega_{\mathrm{l}}.
	\end{align}
	$\gamma_{\mathrm{c}}=(\gamma_{0}+\gamma_{\mathrm{ex}})/2$ is the total optical loss, $\gamma_{0}=\omega_{\mathrm{c}}/Q$ is the intrinsic loss and $\gamma_{\mathrm{ex}}$is the coupling loss. $Q$ is the optical quality factor of the resonator. $i\kappa$ is the dissipative coupling arising from the taper-scattering-induced dissipative backscattering~\cite{Lai2019}. Such a Hamiltonian is  $\mathcal{APT}$ symmetric when $\Delta_{\mathrm{c}}=0$~\cite{Zhang2020a}, with the eigenfrequencies 
	\begin{align}
		\omega_{\pm}=\Delta_{\mathrm{c}}-i\gamma_{\mathrm{c}}\pm\sqrt{\Delta_{\mathrm{sag}}^{2}-\kappa^{2}}.
	\end{align}

	Taking into account the mechanical mode with a resonant frequency $\omega_{\mathrm{m}} $ and an effective mass $m$, the total Hamiltonian of the system is given by	
	\begin{equation}  \label{Hamiltonian}
		\begin{aligned}
			{H}=& H_{0}+H_{\mathrm{int}}+H_{\mathrm{dr}}\text{,}\\
			{H_{0}}=&\frac{p^{2}}{2m}+\frac{1}{2}m\omega_{\mathrm{m}}^{2}x^{2}+\hbar(\Delta_{+}-i\gamma_{\mathrm{c}})a_{\circlearrowright}^{\dagger}a_{\circlearrowright}\\
			&+\hbar(\Delta_{-}-i\gamma_{\mathrm{c}})a_{\circlearrowleft}^{\dagger}a_{\circlearrowleft},\\{H_{\mathrm{int}}}=&i\hbar\kappa(a_{\circlearrowright}^{\dagger}a_{\circlearrowleft}+a_{\circlearrowleft}^{\dagger}a_{\circlearrowright})+\hbar g(a_{\circlearrowright}^{\dagger}a_{\circlearrowright}+a_{\circlearrowleft}^{\dagger}a_{\circlearrowleft})x  \text{,} 	\\
			{H_{\mathrm{dr}}}=&i\hbar\varepsilon_{\mathrm{p}}(a_{\circlearrowright}^{\dagger}e^{-i\xi t}-a_{\circlearrowright}e^{i\xi t})\\
			&+i\hbar\varepsilon_{\mathrm{p}}(a_{\circlearrowleft}^{\dagger}e^{-i\xi t}-a_{\circlearrowleft}e^{i\xi t})\\
			&+i\hbar\varepsilon_{\mathrm{l}}(a_{\circlearrowright}^{\dagger}-a_{\circlearrowright})+i\hbar\varepsilon_{\mathrm{l}}(a_{\circlearrowleft}^{\dagger}-a_{\circlearrowleft}),
		\end{aligned}
	\end{equation}
	where $ x $ and $ p $ are the position and momentum operators of the mechanical mode, $a_{\circlearrowright}$ ($a_{\circlearrowleft}$) and $a_{\circlearrowright}^{\dagger}$ ($a_{\circlearrowleft}^{\dagger}$) denote the annihilation and creation operators of the CW (CCW) mode. $g=\omega_{\mathrm{c}}/R$ is the single-photon coupling rate. The amplitudes of the pump fields at frequency $ \omega_{\mathrm{l}} $ and weak probe field at frequency $\omega_{\mathrm{p}}$ are
	\begin{align}
		\varepsilon_{\mathrm{l}}=\sqrt{\gamma_{\mathrm{ex}}P_{\mathrm{l}}/\hbar\omega_{\mathrm{l}}},\quad \varepsilon_{\mathrm{p}}=\sqrt{\gamma_{\mathrm{ex}}P_{\mathrm{p}}/\hbar\omega_{\mathrm{p}}},
	\end{align}
	where $P_{\mathrm{l}}$ is the pump power and $ P_{\mathrm{p}} $ is the probe power.
	$\xi=\omega_{\mathrm{p}}-\omega_{\mathrm{l}}$ is the detuning between probe and pump field.
	In order to explore the optomechanical effects of the $\mathcal{APT}$-symmetric system, here we set $\Delta_{\mathrm{c}}=\omega_{\mathrm{m}}$~\cite{Carlo2019}.
	The Heisenberg equations of motion (EOM) of this system can be derived from Eq.~(\ref{Hamiltonian}) as
		\begin{equation}
			\begin{aligned}
				\ddot{x} &=-\Gamma_{\mathrm{m}}\dot{x}-\omega_{\mathrm{m}}^{2}x-\frac{\hbar g}{m}(a_{\circlearrowright}^{\dagger}a_{\circlearrowright}+a_{\circlearrowleft}^{\dagger}a_{\circlearrowleft}), \\
				\dot{a}_{\circlearrowright} &=-i(\Delta_{+}-i\gamma_{\mathrm{c}}+gx)a_{\circlearrowright}+\kappa a_{\circlearrowleft}+\varepsilon_{\mathrm{l}}+\varepsilon_{\mathrm{p}}e^{-i\xi t}\text{,}\\
				\dot{a}_{\circlearrowleft}	&=-i(\Delta_{-}-i\gamma_{\mathrm{c}}+gx)a_{\circlearrowleft}+\kappa a_{\circlearrowright}+\varepsilon_{\mathrm{l}}+\ensuremath{\varepsilon_{\mathrm{p}}e^{-i\xi t}},
			\end{aligned}
		\end{equation}
	where $\Gamma_{\mathrm{m}}$ is the mechanical damping rate.
	
	The probe light can be taken as a perturbation for $\varepsilon_{\mathrm{p}}\ll\varepsilon_{\mathrm{l}}$, and thus we expand each operator as the sum of its steady-state value and a small fluctuation around that value, i.e.,
	\begin{align}
		a_{\circlearrowright(\circlearrowleft)}=\bar{a}_{\circlearrowright(\circlearrowleft)}+\delta a_{\circlearrowright(\circlearrowleft)},\quad x=\bar{x}+\delta x,
	\end{align}
	where the steady-state solutions of the system are 
	\begin{equation}
		\begin{aligned}
			\bar{x}	&=\frac{-\hbar g}{m\omega_{\mathrm{m}}^{2}}(\left|\bar{a}_{\circlearrowright}\right|^{2}+\left|\bar{a}_{\circlearrowleft}\right|^{2})\text{,}\\
			\bar{a}_{\circlearrowright}	&=\frac{(i\Delta_{-}+ig\bar{x}+\gamma_{\mathrm{c}}+\kappa)\varepsilon_{\mathrm{l}}}{(i\Delta_{-}+ig\bar{x}+\gamma_{\mathrm{c}})(i\Delta_{+}+ig\bar{x}+\gamma_{\mathrm{c}})-\kappa^{2}},\\
			\bar{a}_{\circlearrowleft} &=\frac{(i\Delta_{+}+ig\bar{x}+\gamma_{\mathrm{c}}+\kappa)\varepsilon_{\mathrm{l}}}{(i\Delta_{-}+ig\bar{x}+\gamma_{\mathrm{c}})(i\Delta_{+}+ig\bar{x}+\gamma_{\mathrm{c}})-\kappa^{2}}.
		\end{aligned}
	\end{equation}
	
	To find the OMIT spectrum of this system, we expand $\delta a_{\circlearrowright(\circlearrowleft)}$ and $\delta x$ by using the following ansatz
	\begin{align}
		\left(\begin{array}{c}
			\delta x\\
			\delta a_{\circlearrowright}\\
			\delta a_{\circlearrowleft}
		\end{array}\right)=\left(\begin{array}{c}
			\delta x_{+}\\
			\delta a_{\circlearrowright+}\\
			\delta a_{\circlearrowleft+}
		\end{array}\right)e^{-i\xi t}+\left(\begin{array}{c}
			\delta x_{-}\\
			\delta a_{\circlearrowright-}\\
			\delta a_{\circlearrowleft-}
		\end{array}\right)e^{i\xi t},
	\end{align}
	and solve the corresponding EOM. Then we can get the solutions as
	\begin{equation}  \label{EqSolutions}
		\begin{aligned}
			\delta a_{\mathrm{\circlearrowright+}}&=\frac{AB_{\circlearrowright}V_{2}+i\hbar g^{2}N_{\circlearrowright}}{AV_{1}V_{2}-i\hbar g^{2}M}\varepsilon_{\mathrm{p}},\\
			\delta a_{\mathrm{\circlearrowleft+}}&=\frac{AB_{\circlearrowleft}V_{2}+i\hbar g^{2}N_{\circlearrowleft}}{AV_{1}V_{2}-i\hbar g^{2}M}\varepsilon_{\mathrm{p}},
		\end{aligned}
	\end{equation}
	where
	\begin{align*}
		&A =m(-\xi^{2}-i\xi\Gamma_{\mathrm{m}}+\omega_{\mathrm{m}}^{2}), \\
		&B_{\circlearrowright(\circlearrowleft)}=i(\ensuremath{\Delta_{-(+)}}-i\gamma_{\mathrm{c}}+g\bar{x}-\xi),\\
		&C_{\circlearrowright(\circlearrowleft)}^{*}=i(\ensuremath{\Delta_{-(+)}}-i\gamma_{\mathrm{c}}+g\bar{x}+\xi),\\ &V_{1}=B_{\circlearrowright}B_{\circlearrowleft}-\kappa^{2},\quad V_{2}=C_{\circlearrowright}C_{\circlearrowleft}-\kappa^{2},\\
	\end{align*}
	and
	\begin{align*}
		N_{\circlearrowright}=&C_{\circlearrowright}B_{\circlearrowright}\left|\bar{a}_{\circlearrowright}\right|^{2}+(C_{\circlearrowleft}B_{\circlearrowright}-V_{2})\left|\bar{a}_{\circlearrowleft}\right|^{2} \\
		&+\kappa B_{\circlearrowright}(\bar{a}_{\circlearrowright}^{*}\bar{a}_{\circlearrowleft}+\bar{a}_{\circlearrowleft}^{*}\bar{a}_{\circlearrowright}), \\
		N_{\circlearrowleft}=&
		C_{\circlearrowleft}B_{\circlearrowleft}\left|\bar{a}_{\circlearrowleft}\right|^{2}+(C_{\circlearrowright}B_{\circlearrowleft}-V_{2})\left|\bar{a}_{\circlearrowright}\right|^{2} \\
		&+\kappa B_{\circlearrowleft}(\bar{a}_{\circlearrowright}^{*}\bar{a}_{\circlearrowleft}+\bar{a}_{\circlearrowleft}^{*}\bar{a}_{\circlearrowright}), \\
		M=&	(B_{\circlearrowright}V_{2}-C_{\circlearrowleft}V_{1})\left|\bar{a}_{\circlearrowright}\right|^{2}+(B_{\circlearrowleft}V_{2}-C_{\circlearrowleft}V_{1})\left|\bar{a}_{\circlearrowleft}\right|^{2} \\
		&+\kappa(V_{2}-V_{1})(\bar{a}_{\circlearrowright}^{*}\bar{a}_{\circlearrowleft}+\bar{a}_{\circlearrowleft}^{*}\bar{a}_{\circlearrowright}).
	\end{align*}
	The solutions in Eq.~(\ref{EqSolutions}) correspond to two different incidence directions of the probe light.
	Then the expectation value of the output field can be derived by using the input-output relation~\cite{gardiner1985}
	\begin{align}
		a_{\circlearrowright(\circlearrowleft)}^{\mathrm{out}}=a_{\circlearrowright(\circlearrowleft)}^{\mathrm{in}}-\sqrt{\gamma_{\mathrm{ex}}} a_{\circlearrowright(\circlearrowleft)},
	\end{align}
	with $a_{\circlearrowright(\circlearrowleft)}^{\mathrm{out}}$ and $a_{\circlearrowright(\circlearrowleft)}^{\mathrm{in}}$ being the output and input field operators.
	The transmission rate of the probe light can be obtained as
	\begin{align}\label{key1}
		\nonumber
		T_{\circlearrowright(\circlearrowleft)}&=\left| t_{\circlearrowright(\circlearrowleft)} \right|^{2}=\left|\frac{a_{\circlearrowright(\circlearrowleft)}^{\mathrm{out}}}{a_{\circlearrowright(\circlearrowleft)}^{\mathrm{in}}}\right|^{2}\\
		&=\left|1-\gamma_{\mathrm{ex}}\frac{AB_{\circlearrowright(\circlearrowleft)}V_{2}+i\hbar g^{2}N_{\circlearrowright(\circlearrowleft)}}{AV_{1}V_{2}-i\hbar g^{2}M}\right|^{2}.
	\end{align}
	
	\begin{figure*}[t]
		\centering
		\includegraphics[width=0.95\linewidth]{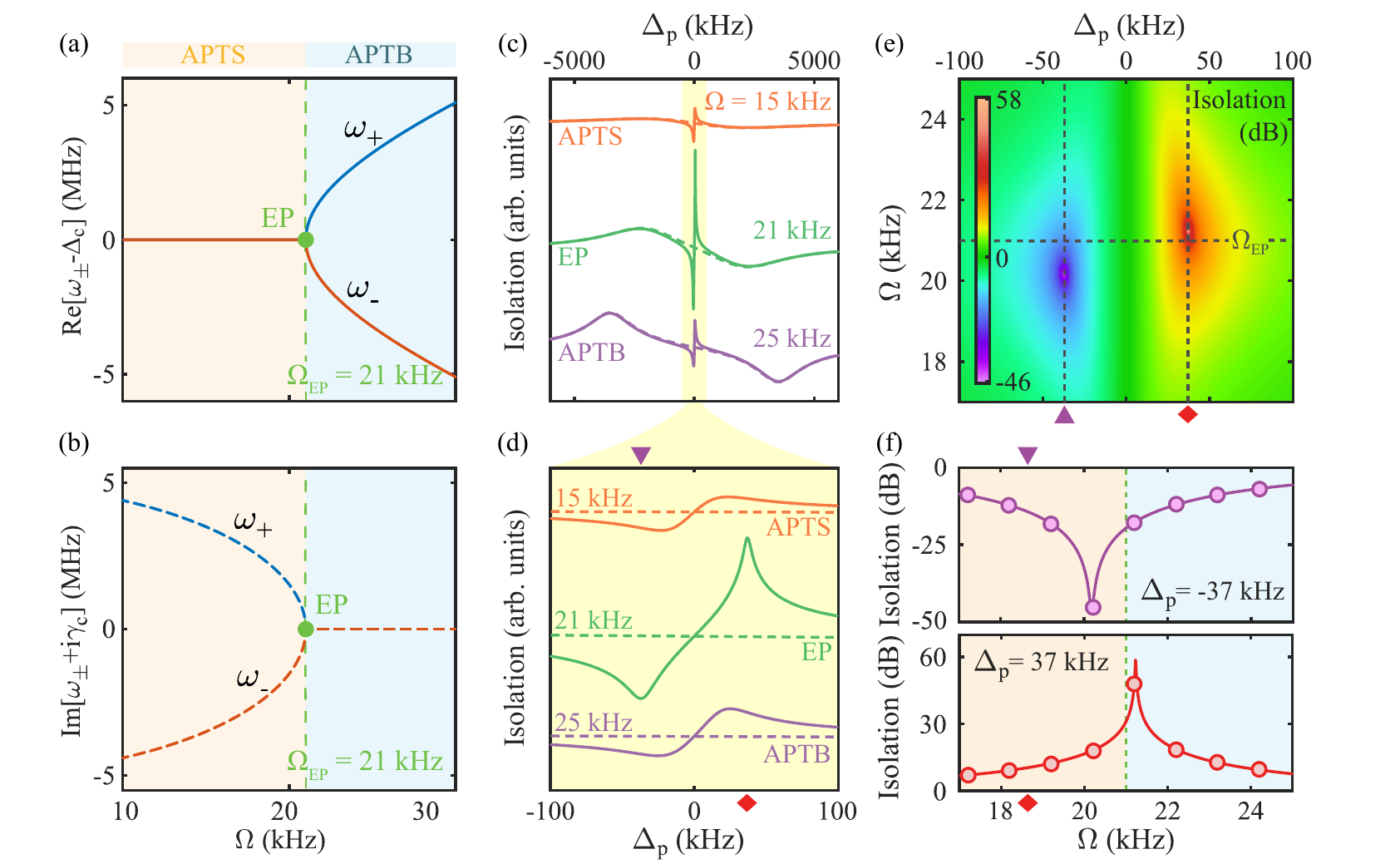}
		\caption{Nonreciprocal light isolation in $\mathcal{APT}$-symmetric ($\mathcal{APTS}$) phase and $\mathcal{APT}$-symmetry-broken ($\mathcal{APTB}$) phase. (a,b) The eigenfrequencies versus the spinning speed $\Omega$. (c,d) Normalized isolation ratio versus optical probe detuning $\Delta_{\mathrm{p}}$ for different spinning speed $\Omega$. The maximum isolation ratio reaches its maximum or minimum around $\Omega= \Omega_{\mathrm{EP}}$.  The colored solid curves indicate the isolation in the presence of pump fields, and the colored dashed curves indicate the isolation in the absence of pump fields. (e) Isolation ratio versus optical probe detuning $\Delta_{\mathrm{p}} $ and spinning speed $\Omega$. (f) For selected $\Delta_{\mathrm{p}}$, the dependence of isolation ratio on spinning speed $\Omega$. The parameters used are the same as those in Fig.~\ref{fig:figure1}(b).\label{fig:figure2} }
	\end{figure*}

	As shown in Ref.~\cite{Zhang2020a}, tuning the spinning speed of a purely optical resonator will not only lead to the phase transition but also nonreciprocal light transmission. In order to explore the similar effect in the optomechanical system, we define an isolation ratio as~\cite{li2022optical}
	\begin{align}\label{Eqisolation}
		I=10\mathrm{log}_{10}\frac{T_{\circlearrowright}}{T_{\circlearrowleft}}.
	\end{align}
	
	This establishes the basis for our discussion of the impact of rotation on the nonreciprocal light isolation and the group delay of the probe light. In Fig.~\ref{fig:figure1}(b), we list the experimental accessible parameters used for numerical simulations in this work.
	
	\section{Nonreciprocal light isolation and group delay or advance}\label{Sec:Results1}
	Figure~\ref{fig:figure2} shows the evolution of eigenfrequencies and nonreciprocal light isolation ratio in $\mathcal{APT}$-symmetric ($\mathcal{APTS}$) phase and $\mathcal{APT}$-symmetry-broken ($\mathcal{APTB}$) phase.	 The evolution of the real and imaginary parts of eigenfrequencies are shown in Figs.~\ref{fig:figure2}(a) and ~\ref{fig:figure2}(b), respectively. When $\Delta_{\mathrm{sag}}$ is increased and becomes $\kappa$, i.e., $\Omega=21\:\mathrm{kHz}$, this system will exhibit an EP at the square-root branch point, i.e., the two eigenstates coalesce. Beyond this point, the system exhibits a transition from an $\mathcal{APTS}$ phase to an $\mathcal{APTB}$ phase, the real part of the frequency of the two states bifurcates and the imaginary part coalesce. In Figs.~\ref{fig:figure2}(c) and ~\ref{fig:figure2}(d), the isolation ratio is plotted as a function of $\Delta_{\mathrm{p}} = \omega_{\mathrm{p}} - \omega_{\mathrm{c}}$. Clearly, nonreciprocal light transmission can be identified in the vicinity of EP.  We note that by tuning the speed to move the system far away from the EP, the influence of the pump fields on the optical isolation tends to be weakened; in contrast, the optical isolation of the system can be significantly enhanced by approaching the EP [see Figs.~\ref{fig:figure2}(c) and (d)].  The isolation ratio is plotted as a function of $\Omega$ and $\Delta_{\mathrm{p}}$ in Fig.~\ref{fig:figure2}(e) to give a comprehensive view.  To see how the isolation ratio varies with rotation speed, in Fig.~\ref{fig:figure2}(f), we plot the isolation ratio as a function of the rotation speed $\Omega$ when the probe detuning is chosen to be $\Delta_{\mathrm{p}}=\pm37\,\mathrm{kHz}$. The isolation ratio of this $\mathcal{APT}$-symmetric optomechanical system reaches its minimum and maximum near the EP, and the maximum is located in the $\mathcal{APTB}$ phase while the minimum is located in the $\mathcal{APTS}$ phase, due to the different probe detunings.

	\begin{figure*}[t]
		\centering
		\includegraphics[width=0.95\linewidth]{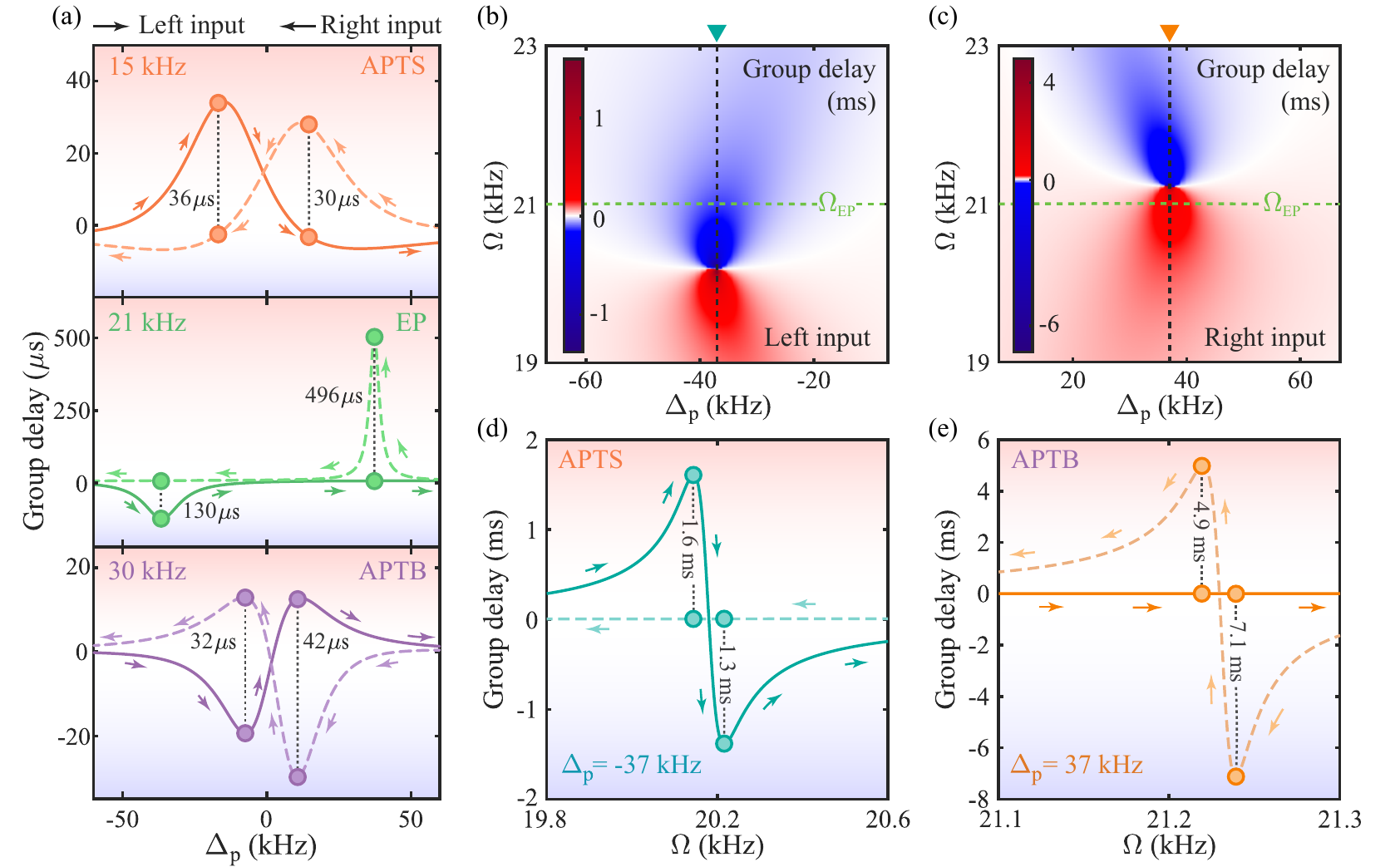}
		\caption{Nonreciprocal group delay. (a) Group delay versus optical probe detuning $ \Delta_{\mathrm{p}} $ for different spinning speed $ \Omega $. The solid (dashed) curves indicate the input from the left (right) hand side. The group delay of the left input field (b) and the right input field (c) versus optical probe detuning $ \Delta_{\mathrm{p}} $ and rotational speed $ \Omega $. (d,e) For selected $ \Delta_{\mathrm{p}} $, the dependence of group delay on spinning speed $ \Omega $. The parameters used are the same as those in Fig.~\ref{fig:figure1}(b). \label{fig:figure3}}
	\end{figure*}

	We stress that the nonreciprocal light transmission in this system is due to the interplay of the optomechanical interaction and the Sagnac effect, which is clearly distinct from the nonreciprocal effects solely originating from the optomechanical interaction~\cite{lu2017,mirza2019optical,Jiang2018,huang2018nonreciprocal,li2019nonreciprocal,li2021nonreciprocal} or the rotation of the resonator~\cite{Maayani2018}.

	Accompanied by the transparency window, the fast or slow light can also be observed, which is characterized by the group delay of the probe light~\cite{SafaviNaeini2011}
	\begin{align}\label{Eqgroupdelay} \tau_{g\circlearrowright(\circlearrowleft)}=\frac{\mathrm{d}\:\mathrm{arg}\left[t_{\circlearrowright(\circlearrowleft)}\right]}{\mathrm{d}\Delta_{\mathrm{p}}}.
	\end{align}
	
	The group delays in various situations are analyzed in detail in Fig.~\ref{fig:figure3}. Figure~\ref{fig:figure3}(a) shows the group delay $\tau_{g\circlearrowright(\circlearrowleft)}$ as a function of $\Delta_{\mathrm{p}}$ for the system varying from $\mathcal{APTS}$ phase to $\mathcal{APTB}$ phase. Similar to the isolation ratio, the group delay of the left or right input probe field also reaches its maximum or minimum in the vicinity of EP. Interestingly, a slow-to-fast light switch can be achieved when increasing the spinning speed to surpass the EP, for both the left and right input fields (but in different regimes of $\Delta_{\mathrm{p}}<0$ or $\Delta_{\mathrm{p}}>0$):
	
	(i) In the $\mathcal{APTS}$ regime, e.g., $\Omega<21\:\mathrm{kHz}$, the group delay $\tau_{g\circlearrowright}$ of the input from left-hand side increased sharply as $\Delta_{\mathrm{p}}$ goes up, reaching its peak in $\Delta_{\mathrm{p}}<0$ regime. Then $\tau_{g\circlearrowright}$ decreases till $\tau_{g\circlearrowright}=0$ at a critical value of $\Delta_{\mathrm{p}}$. Increasing $\Delta_{\mathrm{p}}$ beyond this critical value completes the switch from slow to fast light. After this switch, the advancement of the pulse increases with increasing $\Delta_{\mathrm{p}}$ until it hits its minimum value. And then, beyond this point, increasing $\Delta_{\mathrm{p}}$ pushes $\tau_{g\circlearrowright}$ closer to zero. In contrast, $\tau_{g\circlearrowleft}$ of the input from the right-hand side can experience an opposite trend. 
	
	(ii) In the $\mathcal{APTB}$ regime, increasing the value of $\Delta_{\mathrm{p}}$ pushes the system into the fast light regime and increases the advancement of the pulse until the minimum value. Then the advance effect can be weakened by increasing $\Delta_{\mathrm{p}}$ and finally $\tau_{g\circlearrowright}$ becomes positive, i.e., shifting the fast light to the slow light. If $\Delta_{\mathrm{p}}$ is further increased, $\tau_{g\circlearrowright}$ can be increased significantly till reaching its maximum. For the input signal coming from the right-hand side, an opposite trend is also clearly seen for $\tau_{g\circlearrowleft}$, which is similar to the $\mathcal{APTS}$ regime.

	\begin{figure*}[t]
		\centering
		\includegraphics[width=0.95\linewidth]{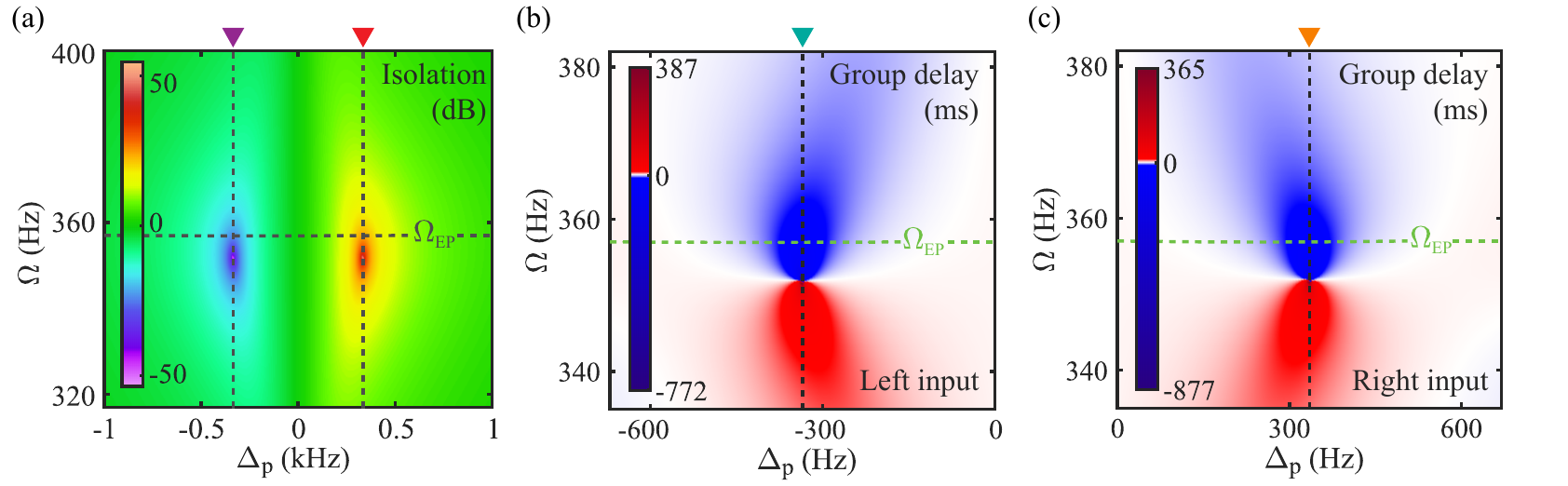}
		\caption{Nonreciprocal light isolation and group delay in $\mathcal{APT}$-symmetric optomechanical systems consisting of an optical microsphere resonator and a nanostring mechanical resonator. (a) Isolation ratio versus optical probe detuning $ \Delta_{\mathrm{p}} $ and rotational speed $ \Omega $. Group delays of the left input field (b) and the right input field (c) versus optical probe detuning $ \Delta_{\mathrm{p}} $ and rotational speed $ \Omega $. Here, we take the experimentally accessible parameters~\cite{armani2003ultra,huet2016millisecond,peng2014loss,schilling2016near,Lai2019}, and $ \Omega_{\mathrm{EP}}=357\:\mathrm{Hz} $.}
		\label{fig:figure4}
	\end{figure*}

	In Figs.~\ref{fig:figure3}(b) and~\ref{fig:figure3}(c), the group delay is plotted as a function of spinning speed $\Omega$ and probe detuning $\Delta_{\mathrm{p}}$. We see that the slow-to-fast light switch of the left input field does not fully coincide with that of the right input field, due to the different Sagnac-Fizeau shifts in the CW and CCW modes induced by the spinning of the resonator. The slow-to-fast light switch is also shown in Figs.~\ref{fig:figure3}(d) and~\ref{fig:figure3}(e), where the probe detuning is set to $\Delta_{\mathrm{p}}=\pm37\:\mathrm{kHz}$. By increasing the value of $\Omega$, the group delay of the left input probe increases significantly, till reaching the maximum value. Beyond the EP, the system enters the fast light regime and the group delay approaches $\tau_{g\circlearrowright}=0$ [see Fig.~\ref{fig:figure3}(d)]. The slow-to-fast light switch of the left input probe field appears in the $\mathcal{APTS}$ phase, while the similar switch of the right input probe appears in the $\mathcal{APTB}$ phase [see Fig.~\ref{fig:figure3}(e)]. 	Clearly, for the spinning optomechanical system, one can not only just tune the system to switch from slow to fast light, or vice versa, by controlling $\Omega$ or $ \Delta_{\mathrm{p}} $, but also achieve nonreciprocal group delay by reversing the direction of the input probe field.
	
	We stress that the observation of these interesting effects is unnecessarily limited to the system as we depicted in Fig.~\ref{fig:figure1}. For example, we can also consider a system consisting of an optical WGM resonator and a nanomechanical beam 100-$\mathrm{\mu m}$-long, 1-$\mathrm{\mu m}$-wide and 0.1-$\mathrm{\mu m}$-thick~\cite{schilling2016near}. The nanomechanical beam with a high resonant frequency, extremely small mass and ultra-high quality factor ($>10^{5}$) coupled to an optical resonator was experimentally realized in previous works~\cite{schilling2016near,gavartin2012hybrid,anetsberger2009near,cole2015evanescent}. For such a system, we choose experimentally available values~\cite{armani2003ultra,huet2016millisecond,peng2014loss,schilling2016near}: i.e., $ \omega_{\mathrm{c}}=193\:\mathrm{THz}$, $ \gamma_{\mathrm{c}}=1.93\:\mathrm{kHz}$, $ P_{\mathrm{l}}=10\:\mathrm{{pW}}$, $ \omega_{\mathrm{m}}=63\:\mathrm{MHz}$, $ \Gamma_{\mathrm{m}}=63\:\mathrm{Hz}$, $ m=10\:\mathrm{pg}$, $ R=50\:\mathrm{\mu m}$, $ g=3.86\:\mathrm{GHz/nm}$. As demonstrated in the experiment~\cite{Lai2019}, the dissipative coupling originating from taper scattering can be about $ 8.5\:\mathrm{kHz}$. We see that similar nonreciprocal light isolation and slow light effects can also be observed in such a system consisting of a $100\times1\times0.1\:\mathrm{\mu m^{3}}$ nanomechanical beam and an optical WGM resonator, as clearly shown in Fig.~\ref{fig:figure4}.
	
	In Fig.~\ref{fig:figure4}(a), the minimum and maximum isolation ratios both occur in the vicinity of $\Omega=\Omega_{\mathrm{EP}}$ and respectively correspond to the probe detuning $\Delta_{\mathrm{p}}=\pm335\:\mathrm{Hz}$. Moreover, the slow-to-fast light switch is still present in the vicinity of $\Omega=\Omega_{\mathrm{EP}}$. This implies that similar performance can be obtained in such an $\mathcal{APT}$-symmetric optomechanical system. The slight difference is that compared with the system with a mechanical radial breathing mode [see Fig.~\ref{fig:figure3}(b)], the maximum and minimum values of group delay of the system with a nanomechanical beam are greatly improved [see Figs.~\ref{fig:figure4}(b) and~\ref{fig:figure4}(c)].

	We note that the OMIT-based slow-to-fast light switch has been extensively studied in different systems, such as  microcavities~\cite{Jing2015,jiao2016nonlinear,asano2016controlling}, circuits~\cite{Zhou2013},atoms~\cite{gu2015tunable,akram2015tunable}, and magnonic systems~\cite{zhao2021phase,lu2021exceptional}. In particular, for a $\mathcal{PT}$-symmetric optomechanical system~\cite{Jing2015}, such a switch was predicted by tuning the gain-loss ratio or pump power. In contrast to these previous efforts~\cite{Jing2015,jiao2016nonlinear,asano2016controlling}, our work here considers an $\mathcal{APT}$-symmetric system, which does not require any gain and thus provides a new route towards realizing such an optical switch using a purely lossy system. In addition, by tuning both the probe detuning $\Delta_{\mathrm{p}}$ and the rotational speed $\Omega$, both nonreciprocal light transmission and nonreciprocal slow or fast light can be achieved by using a single device. In a broader view, the spinning technique can be further combined with other existing methods for the explorations of new effects and applications, such as enhanced light-matter interactions~\cite{ian2008cavity}, on-chip optical storage or processing~\cite{Lake2021} and asymmetric optomechanical entanglement~\cite{liu2023phase}.

	\section{Conclusions}\label{Sec:Conclusions}
	We have investigated the OMIT and the associated group delay in an $\mathcal{APT}$ symmetric optomechanical system. We find that the light propagation becomes highly nonreciprocal by breaking the $\mathcal{APT}$ symmetry, and the isolation ratio reaches its maximum by approaching the EP. Similar to the $\mathcal{PT}$ symmetric system~\cite{Jing2015}, a slow-to-fast light switch can also be identified in our purely lossy system. In particular, we find that the group delay can also become highly nonreciprocal in such a system, enabling the nonreciprocal coexistence of slow and fast light. Hence our work provides a promising new way to manipulate or switch both light transmissions and optical group delay or advance by breaking the $\mathcal{APT}$ symmetry without relying on any active gain.
	
	In our future works, more $\mathcal{APT}$-symmetric optomechanical effects can be explored, such as the role of $\mathcal{APT}$ symmetry breaking in mechanical amplification or phonon laser~\cite{jing2014pt,Zhang2018a}, topological energy transfer~\cite{Xu2016}, OMIT-assisted cooling of mechanical motion~\cite{Ojanen2014}, ultrasensitive force sensing~\cite{Liu2016}, and microwave-over-optical quantum transfer~\cite{afzelius2009multimode,de2008solid}.
	
	\begin{acknowledgments}
		This work is supported by the National Natural Science Foundation of China (Grants No.~11935006 and~12147156 and~12205054), the Science and Technology Innovation Program of Hunan Province (Grant No.~2020RC4047 and~2021RC2078).
	\end{acknowledgments}
	
	\bibliography{APT_REF_V20230217.bib}
	
\end{document}